\documentclass[twocolumn]{revtex4}

\pdfoutput=1

\usepackage{graphicx}
\usepackage{amssymb}
\usepackage{slashed}
\usepackage{amsmath}
\usepackage{url}

\newcommand{\be}{\begin{equation}}
\newcommand{\ee}{\end{equation}}

\newcommand{\Sec}[1]{Sec.~\ref{#1}}
\newcommand{\Fig}[1]{Fig.~\ref{#1}}
\newcommand{\Tab}[1]{Table~\ref{#1}}

\newcommand{\ttbar}{t\bar t}
\newcommand{\lp}{\ell}

\newcommand{\cref}[1]{Chapter \ref{c.#1}}

\def\beq{\begin{equation}} 
\def\eeq{\end{equation}} 
\newcommand{\ba}{\begin{array}}  
\newcommand{\ea}{\end{array}} 
\newcommand{\bea}{\begin{eqnarray}}  
\newcommand{\eea}{\end{eqnarray} }  
\newcommand{\bal}{\begin{align}}
\newcommand{\eal}{\end{align}}   
\def\bi{\begin{itemize}}  
\def\ei{\end{itemize}}  
\def\ben{\begin{enumerate}}  
\def\een{\end{enumerate}}  
\def\beq{\begin{equation}}  
\def\eeq{\end{equation}}  
\def\bc{\begin{center}}
\def\ec{\end{center}} 
 \def\bt{\begin{table}}
\def\et{\end{table}}  
 \def\btb{\begin{tabular}}
\def\etb{\end{tabular}}

\def\cl{{\mathcal L}}

\newcommand{\gev}{\;\mathrm{GeV}}

\def\mass2{mass${}^2$}

\def\pa{\partial}

\newcommand\simlt{\stackrel{<}{{}_\sim}}

\newcommand{\ti}{\tilde}

\def\ov{\overline}

\begin{document}

\title{Unburied Higgs}
\author{Adam Falkowski}
\affiliation{NHETC and Department of Physics and Astronomy, Rutgers University, Piscataway, NJ 08854}
\author{David Krohn}
\affiliation{Department of Physics, Princeton University, Princeton, NJ 08540}
\author{Jessie Shelton}
\affiliation{Department of Physics, Yale University, New Haven, CT 06511}
\author{Arun Thalapillil}
\affiliation{EFI and Department of Physics, University of Chicago, 5640 South Ellis Avenue, Chicago, IL 60637}
\author{Lian-Tao Wang}
\affiliation{Department of Physics, Princeton University, Princeton, NJ 08540}
\date{\today}
\begin{abstract}
Many models of physics beyond the Standard Model yield exotic Higgs decays.  Some of these, particularly those
in which the Higgs  decays to light quarks or gluons, can be very difficult to discover experimentally.  Here we introduce  
a new set of jet substructure techniques designed to search for such a Higgs when its dominant decay is into gluons
via light, uncolored resonances.  We study this scenario in both $V+h $ and $t\bar t +h $ production channels, and find 
both channels lead to discovery at the LHC with $\gtrsim 5\sigma$ at ${\cal L}\sim 100~{\rm fb}^{-1}$.
\end{abstract}
\maketitle
\section{Introduction}

Discovering the Higgs boson is one of the main physics goals of the LHC program.
While collider search strategies have been well developed for the Standard Model (SM) Higgs, the presence of new light degrees of freedom can dramatically alter Higgs phenomenology.  
For instance, in a class of models with an extended Higgs sector, the Higgs can decay via the cascade $h \to 2 a \to 4 X$, 
where $a$ is an on-shell pseudoscalar and $X$ is a SM state to which the pseudoscalar decays.
When this decay dominates, the branching fractions into the  standard discovery channels such as $h\to \gamma \gamma, \tau \tau, b \ov b$ are  very suppressed, and new LHC strategies have to be developed to discover the Higgs.
Previous studies have considered the pseudoscalar decaying into $2b$, $2 \tau$, $2\mu$, and $2\gamma$ \cite{Chang:2006bw}. 
A more challenging case is when the pseudoscalar decays into light hadronic final states, as is predicted in the ``buried Higgs'' model ~\cite{Bellazzini:2009xt} where the dominant decay is $h \to 2a \to 4$ gluons. 
A further motivation to consider this particular decay is that it is less constrained by existing LEP analyses and may allow a the Higgs mass well below 115 GeV \cite{Abbiendi:2002in}.  

Here we will introduce powerful new jet substructure techniques which enable the LHC to discover a Higgs whose dominant decay is to QCD-like jets via a light uncolored resonance.
Specifically, we will consider the decay $h\rightarrow 2a\rightarrow 4g$ where $m_h\sim 80~{\rm GeV}$--$120~{\rm GeV}$ and $a$ is a 
pseudoscalar with $m_a\simlt 10~{\rm GeV}$.
At first sight discovering this Higgs using its dominant decay mode seems hopeless because the dominant Higgs production channels are swamped by overwhelming QCD backgrounds.
To make progress, we follow the strategy pioneered by Ref.~\cite{Butterworth:2008iy} and consider the Higgs in a {\it boosted} regime.  By going to this extreme kinematical limit we are able to substantially reduce the background to our signal.   However, the backgrounds are still considerable, and whereas Ref.~\cite{Butterworth:2008iy} made use of $b$-tagging to push the boosted Higgs into the discovery region, here the situation is more challenging.  
 
Fortunately, these exotic Higgs decays have three features which can distinguish them from QCD backgrounds: (1) $m_a$ furnishes an additional light scale, (2) the Higgs decay is symmetric, as both $a$'s have equal mass, and (3) both the Higgs and the $a$'s are uncolored.  We will find that the key 
to success lies in employing jet substructure tools sensitive to these characteristics.  We will use these tools to devise  a set of cuts 
that allows us to obtain more than $\sim5\sigma$ signal significance at the LHC with $\sqrt{s} = 14$ TeV and ${\cal L}\sim100~{\rm fb}^{-1}$.

This paper is organized as follows.  \Sec{sec:model} discusses a model illustrating the type of exotic Higgs decay we wish to 
investigate.   In \Sec{sec:substr}, we will introduce jet substructure tools designed to find this Higgs, which we will then employ
 in \Sec{sec:analysis} to study the Higgs in the $V+h $ and $t\bar t +h $ production channels.  \Sec{sec:concl} contains our conclusions.

\section{Illustrative Model}
\label{sec:model}

Higgs decays into light jets occur in well-motivated theoretical frameworks.   
In the presence of a light pseudoscalar particle $a$ with cubic couplings to the Higgs,
the Higgs can undergo the cascade decay $h \to 2a \to 4$ partons.  
This cascade was shown to be the generic decay mode in a class of models where the Higgs is a supersymmetric Goldstone boson~\cite{Bellazzini:2009xt}, 
and is also possible in extensions of the MSSM with an additional singlet superfield~\cite{spencer}.  

In the model of \cite{Bellazzini:2009xt} the Higgs boson $h$ has an effective derivative interaction with the pseudoscalar $a$,  
\beq
\cl_{ha^2} \sim \frac{v}{f^2} h (\pa_\mu a)^2  
\eeq
where $v$ is the electroweak scale,  $f$ is the global symmetry breaking scale, and $c$ is a coefficient of order unity. 
As long as $f$ is not much larger than the electroweak scale the decay $h \to 2a$ dominates over the standard $h \to b \bar b$ mode. 
The pseudoscalar is not stable because it has Yukawa couplings to the SM fermions, $i \ti y_\psi a \bar \psi \gamma_5 \psi$. 
The largest Yukawa coupling is to the 3rd generation quarks, while it is suppressed for leptons and lighter quark generations. 
Thus, for $m_a > 2 m_b \sim 10$ GeV the pseudoscalar decays 
almost exclusively into bottom quarks, resulting in the $h \to 4 b$ cascade. 
For $m_a < 2 m_b$ the structure of the
pseudoscalar Yukawa couplings means that the decay into two gluons via a loop of 3rd generation 
quarks dominates over tree level decays to (e.g.) $2\tau$ or $2c$.
The net result is a $h \to 4g$ cascade decay occurring with a $0.8 \sim 0.9$ branching fraction.  
For this decay mode, the current limit on $m_h$ is only 86 GeV assuming the Higgs is produced with the SM cross section \cite{Abbiendi:2002in}
For simplicity and clarity of presentation, in this paper we assume a $100\%$ Higgs branching fraction into four gluons.  

The production of buried Higgses at the LHC proceeds through similar vertices as in the SM.  
We shall assume here that the Higgs couples to the electroweak bosons and the top quarks 
with the same strength as in the SM, although in some models realizing the buried Higgs 
scenario these couplings may again be slightly modified. 

\section{Jet Substructure Tools}
\label{sec:substr}

A buried Higgs is difficult to discover because its decay products are difficult to distinguish from ordinary QCD radiation.  In the case at hand, because $m_a\ll m_h$, the gluons from each $a$ are very collimated, and so an {\it unboosted} buried Higgs will be resolved as two jets. This will be very difficult to distinguish from the enormous backgrounds from QCD radiation.  The extreme kinematic configuration where the Higgs has a large $p_T$, and is thus resolved entirely in one jet, is far more difficult for background processes to mimic. 
In this regime, the two jets from Higgs decay are themselves collimated into a single fat jet with a characteristic substructure.
We will consider two such boosted scenarios,  $pp\rightarrow hW$ (adopting
the basic kinematic cuts of  Ref.~\cite{Butterworth:2008iy}) and $pp\rightarrow ht\bar t$ with a mildly boosted Higgs.

The first step of our analyses is to cluster our events into relatively large jets and identify a candidate boosted Higgs jet.
We then, along the lines of Ref.~\cite{Butterworth:2008iy,Plehn:2009rk}, use a cleaning procedure to remove contamination from
pileup and underlying event
from the jet and place a cut on its mass.  To make further progress we must look to the distinguishing features of the exotic decays.

One characteristic feature of the signal is that the jets from decays of light pseudoscalars $a$
 have small invariant masses, of order $m_a \simlt 10$ GeV.   This is
 clearly independent of the $a$'s $p_T$, while the invariant mass of a QCD jet grows  with $p_T$:
$\sqrt{\langle m_J^2\rangle} \sim  \frac{\bar C \alpha_s}{\pi} p_T R$,   
where $\bar C  = 3 (4/3)$ for gluon (quark) initiated jets  \cite{Ellis:2007ib}.
Because we work in the boosted regime where the $a$'s have a large $p_T$, we expect the bulk of the QCD background subjets to have masses above 10 GeV.    
Thus, requiring that the average mass of the two hardest subjets be small (throughout we will denote
the $i$th hardest subjet as $j_i$),
\[
\ov m \equiv \frac{m(j_1)+m(j_2)}{2} <10\gev,
\]
is an efficient way to separate signal from background. 

The signal events are also distinguished by the symmetry of their decay products: both subjets arise from particles of equal mass. This can be
distinguished by a cut on  {\em mass democracy}:   
\beq
\alpha = \min\left[\frac{m(j_1)}{m(j_2)},\frac{m(j_2)}{m(j_1)}\right]
\eeq  
At the parton level $\alpha = 1$ at leading order, 
%
%
while for the background there is no reason for the QCD radiation to produce democratic jets.   

Finally, signal and background events differ by their color structure \cite{Gallicchio:2010sw}.
For signal events color is only seen very late in the Higgs decay process: neither the Higgs nor the $a$'s carry color charge  
QCD processes therefore only becomes operative only at the scale $\sim 10$ GeV after the pseudoscalars decay into gluons.
By contrast, the background jets are initiated by hard colored particles, which are color-connected to the 
rest of the event; moreover, there is more phase space for QCD radiation.  
Therefore, we expect that the background has more radiation inside the fat jet cone than the signal does.  
We can quantify this intuition using the flow variable   
\beq
\beta = \frac{p_T(j_3)}{p_T(j_1) + p_T(j_2)},
\eeq 
which is motivated by the fact that the signal is unlikely to yield radiation aside from that constituting the two collimated $a$'s. 
We therefore expect the typical value of $\beta$ for background processes to be much larger than for the signal. Before proceeding, we note 
that $\beta$ can be sensitive to very soft radiation, depending on the cut one uses.  Therefore, we employ $\beta$ with 
a threshold: $p_T^{\rm min}$ and set $\beta = 0$ for $p_T(j_3) < p_T^{\rm min}$.   
Other flow variables could be defined to further boost discovery of the buried Higgs. 
In particular, since signal radiates less, a simple cut on the number of subjets above $p_T^{\rm min} \sim 1$ GeV falling inside the fat jet cone adds more discriminating power. However we have not included this cut here because QCD predictions for the number of soft jets are not entirely reliable at the present stage. 
Measuring the number of tracks emanating from the leading subjets could also efficiently separate signal from background \cite{kyle}.  

\section{Analysis}
\label{sec:analysis}
\begin{figure}
\includegraphics[scale=0.45]{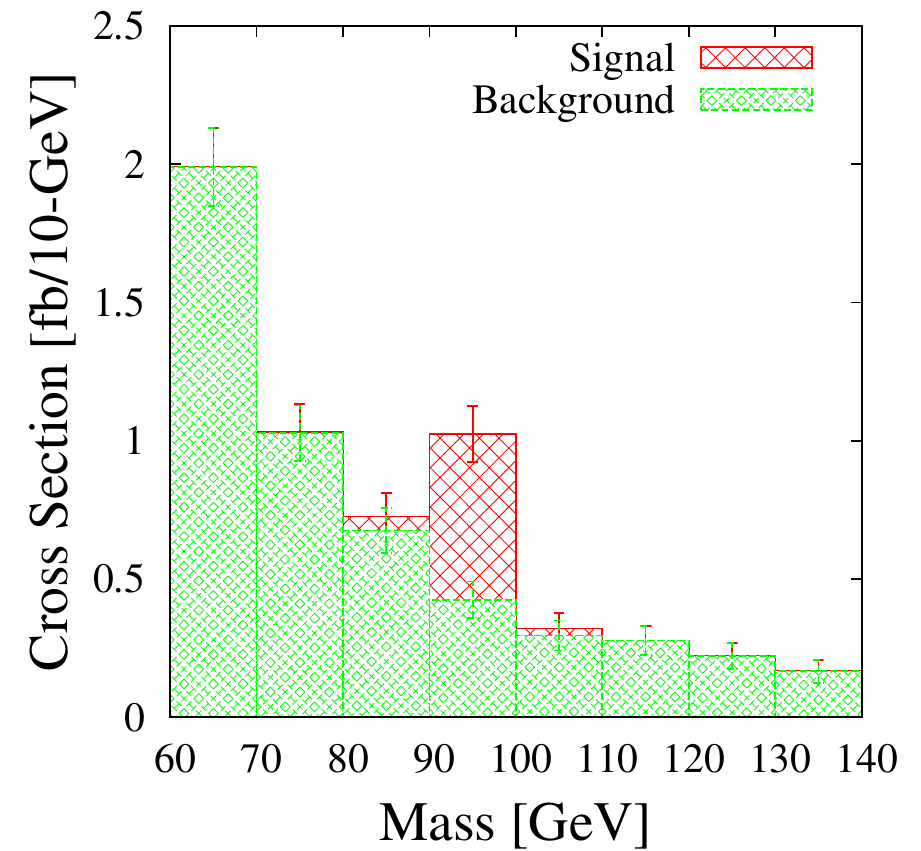}
\includegraphics[scale=0.45]{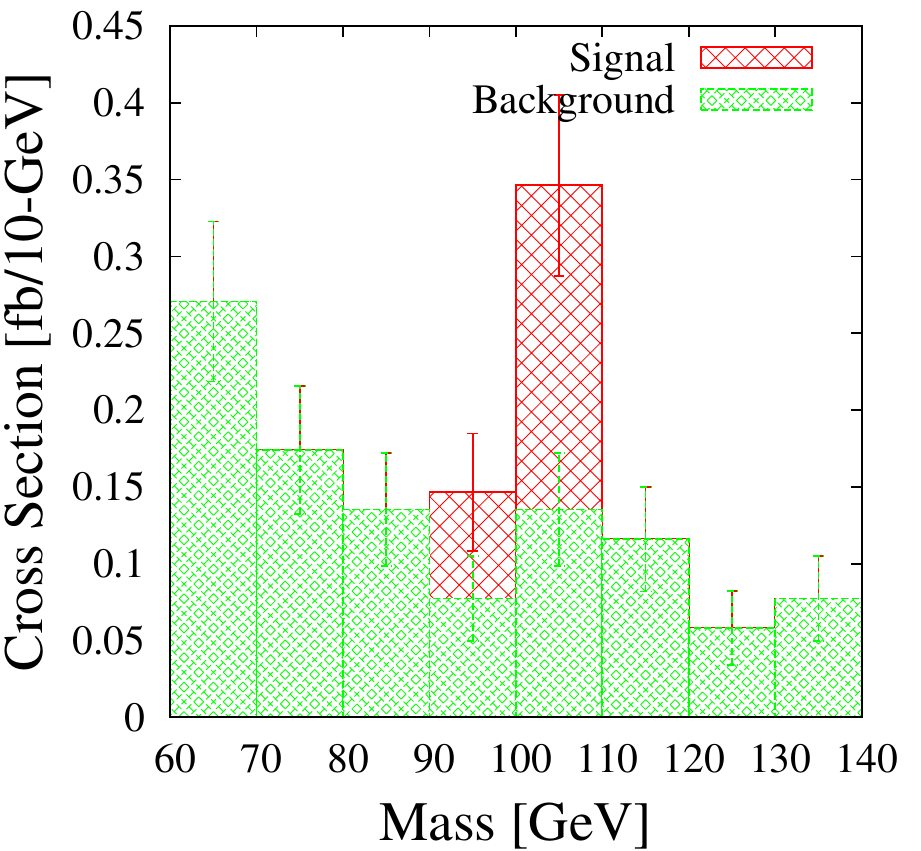}
\caption{\label{fig:reconmass}Reconstructed $m_H=100~{\rm GeV}$ Higgs mass (left) in the $V+h$ channel,
after the cuts of \Tab{tab:cuteffichv} (excluding
the cut on $m_H$); (right) in the $t\bar t+h $ channel, after the cuts of \Tab{table:tthcuts} (excluding the cut on $m_H$).
Error bars show statistical errors.}
\end{figure}
Here we apply the substructure tools developed above to two processes 
yielding a boosted Higgs: $pp\rightarrow hW$ and $pp\rightarrow ht\bar t$.  Before 
proceeding with the analysis we describe our Monte Carlo tools and assumptions.

We generate all signal and background events for $ht\bar t$ at tree level using \texttt{MadGraph v4}~\cite{Maltoni:2002qb} 
and shower them using \texttt{Pythia 6.4.21}~\cite{Sjostrand:2006za}. 
We incorporate underlying event and pile-up
using Pythia's ``DW'' tune and
assuming a luminosity per bunch crossing of $0.05~{\rm mb}^{-1}$.
We generated signal samples for $m_h = 80,100,120$ GeV and $m_a = 8$ GeV. 
Our $\ttbar + $ jets sample is matched out to two
jets using the $k_T$-MLM matching procedure~\cite{Alwall:2007fs} (our $V+$ jets sample 
requires no matching as it is dominated by $2\rightarrow 2$ processes).
Jet clustering is performed using the anti-$k_T$ algorithm~\cite{Cacciari:2008gp} as 
implemented in  \texttt{Fastjet 2.3}~\cite{Cacciari:2005hq}.  When constructing subjets our procedure is to re-cluster the constituents 
of a jet using anti-$k_T$ with a smaller radius, denoted $R_{\rm sub}$.

\subsection{Discovering a buried Higgs in the $V+h$ channel}
Here we consider a boosted Higgs recoiling against a vector boson as in Ref.~\cite{Butterworth:2008iy}.
As the production rate for $pp\rightarrow hW$ is larger than $pp\rightarrow hZ$, and the branching ratio of 
$W$ into leptons is much larger than that of $Z$ into leptons, we will restrict ourselves to the 
process $pp\rightarrow h W$ where $W\rightarrow l\nu$ for $l=e,\mu$.

Our events are clustered 
using jet radii $R$ of 0.8, 1.0, and 1.2 for $m_h$ of 80, 100, and 120 ${\rm GeV}$, respectively.  To force ourselves into the boosted region we will consider events with a jet of $p_T>200~{\rm GeV}$.
The dominant background then is $pp\rightarrow W+j$.  As one can see in \Tab{tab:cuteffichv}, the initial 
backgrounds are horrendous.  Demanding that the average mass
of the hardest two subjets (using $R_{\rm sub}=0.3$) lie below $10~{\rm GeV}$
and requiring the trimmed~\cite{Krohn:2009th} mass of the jet (using the trimming parameter
$f_{\rm cut}=0.03$) lie within $m_h\pm10~{\rm GeV}$ helps, but it is not sufficient for a Higgs discovery.  
 
However, after cutting on the jet substructure variables $\alpha> 0.7$ and $\beta<0.005,0.005,~{\rm and\ }0.007$
 for $m_h$ of 80, 100, and 120 ${\rm GeV}$, respectively, one finds a prominent signal, discoverable regardless of
 whether one uses $p_T^{\rm min}=1~{\rm GeV}$ or a more conservative $5~{\rm GeV}$.
The Higgs mass distribution after
 these cuts is shown in \Fig{fig:reconmass}.
 The final signal significances for 
 the three Higgs masses we consider are shown in \Tab{table:finalsignificance}.  
\label{sec:vh}
\begingroup
\squeezetable
\begin{table}
\caption{Cut efficiencies for a $m_h = 100\gev $ Higgs in the $pp\rightarrow hW$ channel using 
the procedure outlined in \Sec{sec:vh}.  At the end of the table we include results obtained using two different values of  $p_T^{\rm min}$ for $\beta$.\label{tab:cuteffichv}}
\begin{ruledtabular}
\begin{tabular}{c|cccc}
                        & $\sigma_{sig}$ (fb) &  $\sigma_{bg}$ (fb)  & $S/B $  & $S/\sqrt{B}$\\
\hline

$p_T (j) > 200~{\rm GeV}$ & $16$              & $30000$                   &   $0.00052$ & $0.9$     \\

subjet mass              & $12$               & $19000$                   & $0.00062$                    & $0.9$ \\

Higgs window             & $7.1$              & $400$                  & $0.018$                   & $3.6$\\ 

$\alpha > 0.7$            & $4.1 $            &  $140$                    &$0.030 $ & $3.5$ \\
\hline
$\beta < 0.005$, $p_T^{\rm min}=1~{\rm GeV}$       & $0.67$            & $0.74$                    &   $0.90 $ & 7.8 \\
$\beta < 0.005$, $p_T^{\rm min}=5~{\rm GeV}$       & $2.9$                             & $2.6 $                    &   $0.11 $ & 5.7 \\
\end{tabular}

\end{ruledtabular}

\end{table}
\endgroup
\subsection{Discovering a buried Higgs in the $t \bar t+h$ channel}
\label{sec:tth}

Here the signal process of interest is the associated production of a Higgs with a $\ttbar $ pair, followed by leptonic decays of both top quarks and Higgs decaying as  $h \to a a \to 4 g$. 
The final state consists of 2 $b$-tagged jets, 2 opposite-sign leptons, and (at least) 2 hard jets.  
The main background is $\ttbar + $ jets, with secondary contributions from $Z + b\bar b$ and $\ttbar Z$.
Background processes with jets faking a lepton or a $b$-jet are subleading.  For the signal we use the SM NLO $\ttbar H$ cross-section  \cite{Beenakker:2001rj}; in particular $\sigma_{tth} \approx 1$ pb for $m_h = 100$ GeV.  
We use the NLO $+$ NLL calculation of the inclusive $\ttbar + $ jets 
cross-section to normalize the $\ttbar + $ jets background
\cite{Cacciari:2008zb,Dittmaier:2008uj}, $\sigma_{ttj}=908 \;\mathrm{pb} $.  
The NLO cross-section for $\ttbar Z$ is much smaller, $\sigma_{ttZ}=1.1 \;\mathrm{pb} $ \cite{Lazopoulos:2008de}.

Since the buried Higgs does not produce $b$-quarks in its decay, the combinatoric problems that contribute to the difficulty of using the
$t\bar t h$ channel in the SM are significantly ameliorated. In the dileptonic channel, there is in principle no combinatoric background: the decay products of the top quarks can be cleanly separated from the decay products of the Higgs, much as in the $W+h $ channel.  
We first cluster particles using the anti-$k_T $ algorithm with $R_{sub} = 0.4$. 
To select for events containing 2 top quarks decaying leptonically we require two opposite-sign isolated leptons and two $b$-jets satisfying $p_{T,e} > 15\gev$, $p_{T,\mu} >10\gev$, $p_{T, b} > 20\gev$,  $|\eta_{l,b} | <2.5$.  
We assume a flat b-tagging efficiency of $0.6$. 
To control the $Z + b\bar b$ background we require that same-flavor leptons do not reconstruct a $Z$, $ |m_{\lp\lp}-m_Z | >10\gev$.  
After these cuts the cross-section for $Z + b\bar b$ is approximately 10\% of the cross-section for dileptonic $\ttbar +$ jets.  
The importance of $Z + b\bar b$ drops further relative to $\ttbar +$ jets when kinematic cuts are applied, and subsequently we neglect this contribution to the background.

Next, we impose further selection criteria on the remaining untagged jets.
We take jets with $p_T >10\gev $ and further cluster them using the anti-$k_T $ algorithm into fat jets with $R =1.5$.  
We then trim the fat jets by removing the contribution of $R_{sub} = 0.4$ subjets with $p_T < 0.15 \,p_{T,fat}$ from the fat jets.
We select events containing at least one fat jet with $p_T > 125\gev$.
 
The hardest fat jet is our Higgs candidate, and we apply to it similar kinematic and substructure cuts as in the $W+h$ channel.  
We demand that the candidate jet contains at least 2 $R_{sub} = 0.4$ subjets with $p_T > 40\gev$ with the average mass of the hardest two subjets below 10 GeV. Once again, at this stage bump-hunting for a fat jet in the $m_h \pm 10$ GeV mass window is not enough for a discovery, 
and we need to cut on the jet substructure. 
Requiring $\alpha >0.7$ and $\beta  <0.03 $ for $p_{T, min}= 1\gev$ brings us well above the discovery level for $m_h \simlt 100$. 
The cut flow for $m_h =100$ GeV is shown in table~\ref{table:tthcuts}, and the invariant mass distribution of the fat jet mass after all cuts is shown in fig.\ref{fig:reconmass}.     
For $m_h = 120$ GeV we need slightly harder kinematic cuts,  $p_T(j) > 155\gev$, $p_T(j_2) > 50\gev$, $\beta < 0.06$ to lift the significance above the discovery level. 
The final significance for all Higgs masses is given in table~\ref{table:finalsignificance}.

\begingroup
\squeezetable
\begin{table}
\begin{center}
\begin{tabular}{l|cccc}
\hline \hline
                        & $\sigma_{sig}$ (fb) &  $\sigma_{bg}$ (fb) & $S/B $  & $S/\sqrt{B}$ \\
\hline
preselection          & 8.1            & 6700                  & 0.001         & 1.0     \\

$p_T(j) > 125$ GeV & 3.1            & 750                   & 0.004           & 1.1     \\

$p_T(j_2) > 40 \gev,\ov m < 10 \gev$    
 & 0.58        & 22 & 0.03  & 1.2 \\

$m(j) = m_h \pm 10$ GeV      & 0.45           & 3.9                  & 0.1                    & 2.3\\ 

$\alpha > 0.7$            & 0.40           & 2.0                    & 0.2                      & 2.9 \\
\hline
$\beta < 0.03$, $p_T^{min} = 1$ GeV       & 0.28           & 0.21          & 1.3   & 6.1 \\

$\beta < 0.03$, $p_T^{min}= 5$ GeV       & 0.29          & 0.25             & 1.1  & 5.7 \\
\hline \hline
\end{tabular}
\caption{
Cut efficiencies for a $m_h = 100$ GeV Higgs in the $t\bar t h$ channel using the procedure outlined in Sec. IV B.
\label{table:tthcuts}
}
\end{center}

\end{table}
\squeezetable
\begin{table}
\caption{Final signal significance ($S/\sqrt{B}$) and signal-to-background at ${\cal L}=100~{\rm fb}^{-1}$ for three different Higgs masses in the $pp\rightarrow hW$ and $pp\rightarrow ht\bar t$ channels.
The numbers in parenthesis are the significance using $p_T^{\rm min}=5~{\rm GeV}$ for the $\beta$ cut, while those outside the parenthesis are for $p_T^{\rm min}=1~{\rm GeV}$.
\label{table:finalsignificance}
}
\begin{center}
\begin{tabular}{ll|ccc}
\hline \hline
                        && $m_h = 80\gev $ & $m_h = 100\gev $   & $m_h = 120\gev $ \\
\hline
$pp\rightarrow hW$        & $S/\sqrt{B}$ & 6.6 (4.8) &7.8 (5.7) & 7.0 (6.9)  \\
                          & $S/B$        &  0.34 (0.067) & 0.90 (0.11) & 0.80 (0.24) \\
\hline
$pp\rightarrow ht\bar t$  & $S/\sqrt{B}$ & 6.1 (5.9) & 6.1 (5.7) & 7.1 (7.1)  \\
                          & $S/B$ & 1.1 (0.97)  & 1.3 (1.1) & 2.5 (2.5) \\
\hline \hline
\end{tabular}

\end{center}

\end{table}

\endgroup

\section{Conclusions}
\label{sec:concl}

Here we have introduced a set of jet substructure techniques designed to discover a Higgs undergoing  challenging exotic decays. 
Remarkably, we found that these tools  are sufficient to discover a Higgs whose dominant decay is to four gluons in both $W+h$ and $t\bar t+h $ channels after ${\cal L}\sim 100~{\rm fb}^{-1}$.
While the systematic errors in both the background cross sections and the color flow cuts will need to be carefully studied, the comfortable values of $S/B$ which we are able to obtain should ensure that discovery is possible.  
One further lesson is that the $t\bar t+h $ channel can be relatively more useful  for a non-standard Higgs than it is in the SM.
We believe that similar techniques can be applied to boost the LHC discovery potential for a wider class of models where a light Higgs boson undergoes complex decays, e.g. $h \to 4b$ or $h \to 4 \tau$. 

These techniques demonstrate the potential for the LHC to probe qualitatively new scenarios of physics beyond the SM as new jet substructure tools are developed.
One important point of our analysis is that a lot of discriminating power is contained in soft (a few GeV) QCD radiation. 
Further progress in detector sensitivity to soft radiation, as well as a better theoretical control over QCD predictions at the low invariant mass region of the spectrum could lead to further improvement in the discovery potential of non-standard Higgs bosons, or indeed to other non-standard new physics.

{\bf Note added:} When this work was finished Ref. \cite{Chen:2010wk} appeared in which the same Higgs decay is studied with similar conclusions for the LHC discovery potential.

\acknowledgments{We would like to acknowledge useful conversations with Davide Gerbaudo, Pai-hsien Jennifer Hsu, Valery Khoze, Seung Lee, David Miller, Gavin Salam, and Chris Tully.  This work was supported in part by the Yale University Faculty of Arts and Sciences High Performance Computing facility.
The work of AF was supported in part by DOE grant 
DE-FG02-96ER40949.
DK was supported in part by the LHC-TI, and thanks the U. of Washington jet substructure workshop (DOE grant DE-FG02-96ER40956) and the Aspen Center for Physics for their hospitality.  JS was supported in part by DOE grant DE-FG02-92ER40704 and thanks SLAC for its hospitality.
AT was supported in part by the DOE under grant DE-FG02-90ER40560.
LTW is supported by the NSF under grant PHY-0756966 and the DOE under grant DE-FG02-90ER40542.
}


\end{document}